\documentclass[prd,twocolumn,floats,floatfix,nofootinbib,showpacs]{revtex4}
\usepackage{graphicx}
\usepackage{dcolumn}
\usepackage{amssymb}
\usepackage{bm}
\usepackage{amsmath}
\usepackage{color}
\def\spose#1{\hbox to 0pt{#1\hss}}

\def\lta{\mathrel{\spose{\lower 3pt\hbox{$\mathchar"218$}}
     \raise 2.0pt\hbox{$\mathchar"13C$}}}
\def\gta{\mathrel{\spose{\lower 3pt\hbox{$\mathchar"218$}}
     \raise 2.0pt\hbox{$\mathchar"13E$}}}
\newcommand{\be}{\begin{equation}}
\newcommand{\en}{\end{equation}}
\newcommand{\bea}{\begin{eqnarray}}
\newcommand{\ena}{\end{eqnarray}}

\newcommand{\etal}{\textsl{et al.~}}

\begin{document}
\title{An Inflationary Non-singular Quantum Cosmological Model}

\author{Felipe T. Falciano}
\email{ftovar@cbpf.br} \affiliation{Instituto de Cosmologia Relatividade e Astrofisica ICRA -- CBPF, \\ Rua Xavier Sigaud, 150, Urca,
 22290-180, Rio de Janeiro, Brazil}

\author{Nelson Pinto-Neto}
\email{nelsonpn@cbpf.br} \affiliation{Instituto de Cosmologia Relatividade e Astrofisica ICRA -- CBPF, \\ Rua Xavier Sigaud, 150, Urca,
 22290-180, Rio de Janeiro, Brazil}

\author{E. Sergio Santini}
\email{santini@cbpf.br} \affiliation{Instituto de Cosmologia Relatividade e Astrofisica ICRA -- CBPF, \\ Rua Xavier Sigaud, 150, Urca,
 22290-180, Rio de Janeiro, Brazil}
\affiliation{ Comiss\~ao Nacional de Energia Nuclear \\
Rua General Severiano 90, Botafogo 22290-901,  Rio de Janeiro, Brazil}
\date{\today}

\begin{abstract}
A stiff matter-dominated universe modeled by a free massless scalar field minimally coupled to gravity in a Friedmann-Lema\^{\i}tre-Robertson-Walker (FLRW) geometry is quantized. 
Generalized complex-width gaussian superpositions of the solutions of the Wheeler-DeWitt equation are 
constructed and the Bohm-de Broglie interpretation of quantum cosmology is applied. A planar dynamical system 
is found in which a diversity of quantum bohmian trajectories are obtained and discussed. One class of solutions represents non-singular inflationary models starting at infinity past from flat space-time with Planckian size spacelike hypersurfaces, which inflates without inflaton but due to a quantum cosmological effect, until it makes an analytical graceful exit from this inflationary epoch to a decelerated classical stiff matter expansion phase.
\end{abstract}


\maketitle

\section{Introduction}

For more than 25 years, inflation~\cite{inflation} has been considered a
paradigm to solve, at the same time, standard cosmological puzzles
related to initial conditions like the
flatness, horizon, and isotropy problems, and,
as a bonus, astroparticle issues like the monopole excess. More important, it also predicts
that primordial fluctuations, assumed to be of quantum origin,
could be enhanced to the level required to trigger large scale
structure formation, with an almost scale-invariant spectrum \cite{muk},
which is confirmed by observations \cite{wmap}. 

The inflation paradigm is also endowed with two specific problems,
that may ultimately be related, namely
the meaning of the trans-Planckian~\cite{transP} perturbations and the
existence of a past singularity~\cite{singularity}. Concerning the
latter, the existence of an initial singularity is one of the major drawbacks of
classical cosmology. In spite of the fact that the standard cosmological
model, based in the classical general relativity theory, has been
successfully tested until the nucleosynthesis era (around $t \sim 1 s$),
the extrapolation of this model to higher energies leads to a breakdown
of the geometry in a finite cosmic time. This breakdown of the geometry
may indicate that the classical theory must be replaced by a quantum
theory of gravitation: quantum effects may avoid the presence of the
singularity, leading to a complete regular cosmological model.

Among the fundamental questions that come from the quantization of the
universe as a whole, one of the most important concerns the
interpretation of the wave function of the Universe. In order to extract 
predictions from it, the Bohm-de Broglie (BdB) ontological interpretation of quantum
mechanics \cite{bohm, holland} has been proposed
\cite{santini0, cons, bola}, since it avoids many conceptual difficulties
that follow from the application of the standard Copenhagen
interpretation to a unique system that contains everything. In
opposition to the latter one, the ontological interpretation does not
need a classical domain outside the quantized system to generate the
physical facts out of potentialities (the facts are there {\it ab
initio} because the positions and trajectories of the particles (called
bohmian trajectories) are considered
to be part of  objective reality, and hence it can be applied to the Universe as a
whole. There are other alternative interpretations which can be used in quantum
cosmology, like the many worlds interpretation of quantum mechanics \cite{many},
but we will not consider them here because they are probabilistic 
interpretations in essence. As we know \cite{kuc}, it is very
difficult to obtain from the the Wheeler-DeWitt equation, when applied
to a closed universe, a probabilistic interpretation
for their solutions because of its hyperbolic nature (see however further approachs
\cite{har}). In the case of the
Bohm-de Broglie interpretation, probabilities are useful but not essential,
as long as objective trajectories (universe histories) can be calculated and their properties studied.
Probabilities can be recovered, as it
has been suggested many times \cite{banks,pad,kie,hal}, at
the semiclassical level, where a probability measure can be constructed with
the quantum solutions. Hence, we can take the Wheeler-DeWitt equation as it is, without
imposing any probabilistic interpretation at the most fundamental level,
but still obtaining information using the Bohm-de Broglie interpretation,
and then recover probabilities when we reach the semiclassical level.
With this interpretation in hands, one can ask if the
quantum scenario predicted by the wave function of the Universe is free of
singularities, and which type of classical universe emerges from the
quantum phase.

Quantum cosmology, in this framework, exhibits bouncing solutions \cite{bounce} which can be
interpreted as truly avoiding the singularity. Some of these bouncing models provide solutions to 
many cosmological puzzles, and may also yield a scale invariant cosmological
perturbations as in inflationary models \cite{emanuel}. Bouncing models are also
obtained in loop quantum cosmology \cite{loop}.

Some types of bouncing models were obtained in \cite{fab2,santini}. 
In these references, the matter content of the primordial model was
considered to be stiff matter \footnote{A perfect fluid with a ``super-rigid'' equation of state $p=\rho$, that was proposed long time ago in  \cite{zeldovich}.}, modeled by a free massless scalar field minimally
coupled to gravity. The Wheeler-DeWitt equation turns out to be a
two-dimensional Klein-Gordon equation, and gaussian superpositions, 
with positive width, of negative and
positive frequency modes solutions were considered. In the flat case, a
two dimensional dynamical system for the bohmian trajectories was obtained, yielding
a variety of possibilities: big bang-big crunch models, oscillating universes, 
bouncing solutions, and ever expanding (contracting) big bang (big crunch) models with a period of acceleration
in the middle of their evolution. No non-singular purely expanding model with a primordial
accelerated phase (a non-singular inflationary model) was obtained (models of this type
were obtained only with a non-minimal coupling between the scalar and gravitational fields,
\cite{nonminimal}).

In the present paper we generalize the gaussians superpositions of Refs.~\cite{fab2,santini}
to gaussians with complex widths and non negative real part. We obtain a richer
two dimensional dynamical system. When the real part of the width is made zero, we obtain non-singular inflationary models which expand accelerately in the infinity past from 
flat universes with finite (not zero) size spatial sections, and is smoothly connected to a classical decelerated
stiff matter expanding phase. It has features of the pre-big bang model \cite{pbb}
and the emergent model \cite{ellis} without gracefull exit problem.

The paper is organized as follows: in the next section the classical model is presented.
Section III is devoted to its quantization and the corresponding Wheeler-DeWitt equation 
is obtained. Generalized gaussian
superpositions of the quantum solutions, and their
corresponding dynamical system are studied in section IV. In Section V, the non-singular inflationary model
is studied and its properties are presented and discussed. In section VI we present our conclusions.

\section{The classical minisuperspace model}

The model we take contains a massless free scalar field (stiff matter),
and the total lagrangian reads
\begin{equation}
\label{lg2}
{\it L} = \sqrt{-g}\biggr[\frac{R}{6l^2} - \frac{1}{2}\phi_{;\mu}
\phi^{;\mu}\biggl] \quad,
\end{equation}
where we are using natural units $\hbar =c=1$, and $l^2\equiv8\pi G/3$, which is the Planck
length squared in these units. 

We will consider the spatially homogeneous and isotropic space-time line element,
\begin{equation}
\label{m}
ds^2 = -N^2 {\rm d}t^2 + \frac{{a(t)}^2}{(1 + \epsilon r^2/4)^2}[{\rm
d}r^2 + r^2({\rm d} \theta ^2 + \sin ^2 (\theta) {\rm d} \varphi ^2)] \quad,
\end{equation}
where the spatial curvature $\epsilon$ takes the values $0$, $1$,$-1$.
Inserting this line element into the lagrangian (\ref{lg2}), and omitting
a total time derivative, we obtain the following minisuperspace action:
\begin{equation}
\label{lq}
S = \int\biggr(\frac{-{\dot a}^2aV}{Nl^2} + \frac{N\epsilon aV}{l^2}
+ \frac{{\dot\phi}^2a^3 V}{2N}\biggl){\rm d}t \quad,
\end{equation}
where $V$ is the total volume divided by $a^3$ of the spacelike
hypersurfaces, which are supposed to be closed. 
$V$ depends on the value of $\epsilon$ and on the topology of
the hypersurfaces. For $\epsilon =0$, $V$ can have any value because the
fundamental polyhedra of $\epsilon =0$ hypersurfaces can have arbitrary
size (see Ref. \cite{top}). In the case of $\epsilon = 1$ and topology
$S^3$, $V=2\pi ^2$. 

Usually, the scale factor has dimensions of length because we use
angular coordinates in closed spaces. Also, in natural units, the scalar
field has dimensions of the inverse of a length. Hence we will define 
the dimensioless quantities ${\bar a} \equiv \sqrt{2V} a/l$, 
${\bar\phi} \equiv l \phi/\sqrt{2}$. 
Calculating the hamiltonian, and omitting the bars, yields,
\begin{equation}
H = \frac{\sqrt{2V}N}{l}\biggr(-\frac{p_a^2}{2a} + \frac{p_{\phi}^2}
{2a^3} - \frac{\epsilon a}{2}\biggl) \quad.
\end{equation}
As $\sqrt{2V}/l$ appears as an overall multiplicative constant in the
hamiltonian, we can set it equal to one without any loss of generality,
keeping in mind that the physical scale factor which appears in the metric is
$l a/\sqrt{2}$, not $a$. We can further simplify the hamiltonian by defining 
$\alpha \equiv \ln(a)$ obtaining
\begin{equation}
\label{hamalphaphi}
H = \frac{N}{2\exp(3\alpha)} \biggr[-p_\alpha ^2 + 
p_{\phi}^2 - \epsilon \exp(4\alpha) \biggl] \quad,
\end{equation}
where
\begin{eqnarray}
\label{palpha}
p_\alpha &=& -\frac{e^{3\alpha}\dot \alpha}{N} \quad, \\
\label{pphi}
p_\phi &=& \frac{e^{3\alpha}\dot\phi}{N} \quad.
\end{eqnarray}
The momentum $p_\phi$ is a constant of motion which we will call ${\bar k}$. 

The classical solutions are, in the gauge $N=1$ (cosmic time):
\vspace{0.5cm}

{\bf 1) For $\epsilon =0$:}

\begin{equation}
\phi = \pm \alpha + c_1 \quad,
\end{equation}
where $c_1$ is an integration constant. In terms of cosmic time $\tau$ they read:
\begin{eqnarray}
a &=& e^{\alpha} = 3 {\bar k} \tau^{1/3} \quad, \\
\phi &=& \frac{\ln (\tau)}{3} + c_2 \quad.
\end{eqnarray}
The solutions contract or expand forever from a singularity, depending
on the sign of ${\bar k}$, without any inflationary epoch.
\vspace{0.5cm}

{\bf 2) For $\epsilon =1$:}

\begin{equation}
a = e^{\alpha} = \frac{\bar k}{\cosh(2\phi - c_1)} ,
\end{equation}
where $c_1$ is an integration constant, and from the conservation of $p_\phi$
we get
\begin{equation}
{\bar k} = e^{3\alpha}\dot\phi .
\end{equation}
The cosmic time dependence is complicated and we will not write it here.
These solutions describe universes expanding from a singularity till a
maximum size and contracting again to a big crunch. Near the
singularity, these solutions behave as in the flat case. There is no
inflation.
\vspace{0.5cm}

{\bf 3) For $\epsilon =-1$:} 

\begin{equation}
\label{-1}
a = e^{\alpha} = \frac{\bar k}{\mid\sinh(2\phi - c_1)\mid} \quad,
\end{equation}
where $c_1$ is an integration constant, and again, from the conservation of 
$p_\phi$ we get
\begin{equation}
{\bar k} = e^{3\alpha}\dot\phi \quad.
\end{equation}
As before, the cosmic time dependence is complicated and we will not
write it here. These solutions describe universes contracting forever to
or expanding forever from a singularity. Near the singularity, these
solutions behave as in the flat case. There is no inflation.

Hence, in all models there is at least one singularity and no acceleration
phase, as it should be for a classical stiff matter fluid.

\section{Quantization and the Bohm-de Broglie  interpretation}

Let us now quantize the model. The Wheeler-DeWitt equation is obtained
through the Dirac quantization procedure, where the wave function must be
annihilated by the operator version of the Hamiltonian constraint.
For the case of homogeneous minisuperspace models, which have a finite number
of degrees of freedom, the minisuperspace Wheeler-De Witt equation reads
\begin{equation} 
\label{bsc0}
{\cal H}({\hat{p}}^{\mu}, {\hat{q}}_{\mu}) \Psi (q) = 0 \quad.
\end{equation}
The quantities ${\hat{p}}^{\mu}, {\hat{q}}_{\mu}$ are the phase space operators
related to the homogeneous degrees of freedom of the model.
Usually this equation can be written as

\begin{equation}
\label{bsc}
-\frac{1}{2}f_{\rho\sigma}(q_{\mu})\frac{\partial \Psi (q)}{\partial q_{\rho}\partial q_{\sigma}}
+ U(q_{\mu})\Psi (q) = 0 \quad,
\end{equation}
where $f_{\rho\sigma}(q_{\mu})$ is the minisuperspace DeWitt metric of the model, whose inverse is
denoted by $f^{\rho\sigma}(q_{\mu})$.

Writing $\Psi$ in polar form, $\Psi = R \exp (iS)$, and substituting it into (\ref{bsc}),
we obtain the following equations:

\begin{equation}
\label{hoqg}
\frac{1}{2}f_{\rho\sigma}(q_{\mu})\frac{\partial S}{\partial q_{\rho}}
\frac{\partial S}{\partial q_{\sigma}}+ U(q_{\mu}) + Q(q_{\mu}) = 0 \quad,
\end{equation}
\begin{equation}
\label{hoqg2}
f_{\rho\sigma}(q_{\mu})\frac{\partial}{\partial q_{\rho}}
\biggl(R^2\frac{\partial S}{\partial q_{\sigma}}\biggr) = 0 \quad,
\end{equation}
where 

\begin{equation}
\label{hqgqp}
Q(q_{\mu}) \equiv -\frac{1}{2R} f_{\rho\sigma}\frac{\partial ^2 R}
{\partial q_{\rho} \partial q_{\sigma}} 
\end{equation}
is called the quantum potential.

The Bohm -de Broglie interpretation applied to Quantum Cosmology states that the trajectories $q_{\mu}(t)$ are real, independently of any observations. Equation (\ref{hoqg}) represents their Hamilton-Jacobi equation, which is the classical one  added with a quantum potential term Eq.(\ref{hqgqp}) responsible for the quantum effects. This suggests to define

\begin{equation}
\label{h}
p^{\rho} = \frac{\partial S}{\partial q_{\rho}} ,
\end{equation}
where the momenta are related to the velocities in the usual way:

\begin{equation}
\label{h2}
p^{\rho} = f^{\rho\sigma}\frac{1}{N}\frac{\partial q_{\sigma}}{\partial t} .
\end{equation}

To obtain the quantum trajectories we have to solve the following
system of first order differential equations, called the guidance relations:
\begin{equation}
\label{h3}
\frac{\partial S(q_{\rho})}{\partial q_{\rho}} =
f^{\rho\sigma}\frac{1}{N}\dot{q}_{\sigma} .
\end{equation} 

Eqs.(\ref{h3}) are invariant under time reparametrization.
 Hence, even at the quantum level, different choices of $N(t)$ yield the same
space-time geometry for a given non-classical solution $q_{\alpha}(t)$.

There is no problem of time in the Bohm-de Broglie interpretation of
minisuperspace quantum cosmology \cite{bola27}. 
This is not the case, however, for the full superspace, see \cite{santini0}\cite{tese}, 
although the theory remains consistent, see \cite{tese}\cite{cons}).

Let us then apply this interpretation to our minisuperspace model.
The operator version of Eq.~(\ref{hamalphaphi}), with the factor ordering which makes it
covariant through field redefinitions, reads 
\begin{equation}
\label{wdw}
\frac{1}{2e^{3\alpha}}\biggl( -\frac{\partial ^2\Psi}{\partial \alpha ^2} +  \frac{\partial
^2\Psi}{\partial \phi ^2} + \epsilon e^{4\alpha}\Psi\biggr) = 0 \quad,
\end{equation}
from where one can read that $f_{\rho\sigma} = \eta_{\rho\sigma} e^{-3\alpha}$, and $\eta_{\rho\sigma}$
is the Minkowski metric in two dimensions. Comparing Eq.~(\ref{wdw}) with Eqs.~(\ref{bsc}) and (\ref{hoqg})
yields for the quantum potential 
\begin{equation}
Q(\alpha ,\phi )\equiv -\frac{e^{3\alpha}}{R} f_{\rho\sigma}\frac{\partial ^2 R}
{\partial q_{\rho} \partial q_{\sigma}} = \frac{1}{R}\biggr[\frac{\partial^{2}R}
{\partial \alpha^{2}}-\frac{\partial^{2}R}{\partial \phi^{2}}\biggl]\quad .
\end{equation}
The guidance relations (\ref{h3}) read
\begin{equation}
\label{guialpha}
\frac{\partial S}{\partial \alpha}=-\frac{e^{3\alpha}\dot{\alpha}}{N}\quad ,
\end{equation}
\begin{equation}
\label{guiphi}
\frac{\partial S}{\partial \phi}=\frac{e^{3\alpha}\dot{\phi}}{N}\quad .
\end{equation}

We can write this equation in null coordinates,
\begin{eqnarray}
\label{nulas}
u\equiv\frac{1}{\sqrt{2}}(\alpha+\phi) & & \alpha\equiv\frac{1}{\sqrt{2}} \left(u+v\right)\nonumber\\
v\equiv\frac{1}{\sqrt{2}}(\alpha-\phi)  & & \phi\equiv\frac{1}{\sqrt{2}} \left(u-v\right)
\end{eqnarray}
yielding,
\begin{equation}
\left(-\frac{\partial^{2} }{\partial u \partial v }+\frac{\epsilon }{2} e^{2\sqrt{2} \left(u+v\right)} \right) \Psi \left(u,v \right) =0 \quad.
\end{equation}
The solutions are:
\vspace{0.5cm}

{\bf 1) For $\epsilon =0$:}

In this case the general solution is
\begin{equation}
\label{sol0}
\Psi(u,v) = F(u) + G(v) \quad,
\end{equation}
where $F$ and $G$ are arbitrary functions.
Using a separation of variable method, one can write these solutions
as Fourier transforms given by
\begin{equation}
\label{sol0k}
\Psi(u,v) = \int d k U(k)\exp (iku) + \int d k V(k)\exp (ikv) \quad,
\end{equation}
$U$ and  $V$ also being arbitrary.

Writing the solution (\ref{sol0}) in polar form, 
\[
\Psi \left(u,v \right)=R_{+}e^{i S_{+}} +R_{-}e^{i S_{-}}
\]
where
\begin{eqnarray*}
R_{+}=R(u)&&S_{+}=S(u)\\
R_{-}=R(v)&&S_{-}=S(v)\\ \quad,
\end{eqnarray*}
one obtains
\begin{eqnarray*}
\emph{R}&=&\sqrt{R_{+}^{2}+R_{-}^{2}+2R_{+} R_{-}\cos(S_{+}-S_{-})}\\
\emph{S}&=&{\rm arctan}\left(\frac{ R_{+}\sin(S_{+})+R_{-}\sin(S_{-}) }{ R_{+}\cos(S_{+})+R_{-}\cos(S_{-})  }\right)
\end{eqnarray*}
The derivative of $S$ with respect to some variable $x$ reads
\begin{widetext}
\begin{eqnarray*}
\label{flu}
\frac{\partial \emph{S} }{\partial x }&=&\frac{
R^{2}_{+}\frac{\partial S_{+}}{\partial x }+R^{2}_{-}\frac{\partial S_{-}}{\partial x }+\left(\frac{\partial S_{+}}{\partial x }+\frac{\partial S_{-}}{\partial x } \right)R_{+}R_{-}\cos\left(S_{+}-S_{-}\right)+\left(R_{-}\frac{\partial R_{+}}{\partial x }-R_{+}\frac{\partial R_{-}}{\partial x }\right)\sin\left(S_{+}-S_{-}\right)}{R_{+}^{2}+R_{-}^{2}+2R_{+} R_{-}\cos(S_{+}-S_{-})}
\end{eqnarray*}
These equations will be used in the next section.
\end{widetext}
\vspace{0.5cm}

{\bf 2) For $\epsilon \neq 0$:}

In this case the general solution reads
\begin{eqnarray}
\label{sol1}
&&\Psi(u,v) = \int d k\, U(k)\exp \biggl(\frac{k}{2\sqrt{2}}e^{2\sqrt{2}u} + \frac{\epsilon}{4\sqrt{2}k}e^{2\sqrt{2}v}\biggr) \nonumber \\
&+& \int d k\, V(k)\exp \biggl(\frac{k}{2\sqrt{2}}e^{2\sqrt{2}v} + \frac{\epsilon}{4\sqrt{2}k}e^{2\sqrt{2}u}\biggr) \quad ,
\end{eqnarray}
where $f$ and $g$ are arbitrary functions.
In Ref.\cite{fab2} these solutions were expanded in terms of Bessel functions.

\section{Generalized gaussian superpositions}

In Ref.~\cite{fab2, santini} we made gaussian superpositions of the solutions with
the choice $U(k)=V(\pm k) = A(k)$, with $A(k)$ given by
\begin{equation} \label{gauss1}
A(k) = \exp\biggr[-\frac{(k-\sqrt{2}d)^2}{\sigma^2}\biggl] \quad,
\end{equation}
with $\sigma^2>0$, and the presence of $\sqrt{2}$ above is just for further convenience. 
Bouncing non-singular solutions were obtained in \cite{fab2}, and expanding 
singular models with a period of acceleration between decelerated phases were proposed in \cite{santini}.

In this paper we will consider the more general case where the parameter 
$\sigma^2$ in (\ref{gauss1})
is given by a complex number: $\sigma^2+i4h$, where $h$ is an arbitrary real number. 
Under this asumption we have:

 \begin{equation} \label{gauss}
A(k) = \exp\biggr[-\frac{(k-\sqrt{2}d)^2}{\sigma^2+i4h}\biggl] \quad,
\end{equation}
From now on we will consider only flat spatial sections. 

Integrating (\ref{sol0k})  with $U(k)=V(k) = A(k)$, we obtain the solution:

\begin{widetext}
\begin{eqnarray}
\label{felipeG}
\Psi(u,v) &=& \sqrt{\pi}\; \sqrt[4]{\sigma^4+16 h^2}\; e^{i\arctan{\sqrt{\frac{\sqrt{\sigma^4+16 h^2}-\sigma^2}
{\sqrt{\sigma^4+16 h^2}+\sigma^2}}}} \biggl\{\exp \biggl[-\frac{\sigma^2}{4}u^2+i(-h u^2 + \sqrt{2}du) 
\biggr]\nonumber \\ &+&
\exp \biggl[-\frac{\sigma^2}{4}v^2+i(-h v^2 + \sqrt{2}dv )\biggr]\biggr\} \quad.
\end{eqnarray}

To obtain the quantum trajectories, we have to calculate the phase $S$ of
the above wave function
and substitute it into the guidance equations.
Computing the phase and recovering the original variables $\alpha, \phi$, we have 
\begin{equation}
S= \arctan\biggl(\sqrt{\frac{\sqrt{\sigma^4+16 h^2}-\sigma^2}{\sqrt{\sigma^4+16 h^2}+\sigma^2}}\biggr)+ d\alpha-\frac{h}{2}(\alpha^2+\phi^2)+\arctan\biggl\{\tanh\biggl(\frac{\sigma^2 \alpha \phi}{4}\biggr)
\tan[\phi(h \alpha -d)]\biggr\} ,
\end{equation}
which, after substitution in Eqs (\ref{guialpha},\ref{guiphi}), yields a planar system given by:

\begin{eqnarray}\label{gacomplex}
\dot{\alpha}=
-\frac{N}{4e^{3\alpha  }}\biggl\{4(d-h\alpha)+\frac {\sigma^2 \phi \sin [2\phi(h\alpha - d)] + 4h
\phi \sinh \left(\frac{\sigma^2\phi\alpha}{2}\right)}{\cosh \left(\frac{\sigma^2\phi \alpha}{2}  
 \right) +\cos [2\phi(h\alpha - d)]}\biggr\} =: f(\alpha,\phi) ,
\end{eqnarray}
and
\begin{eqnarray}\label{gpcomplex}
\dot{\phi}=\frac{N}{4 e^{3\alpha}} \biggl\{-4h\phi + \frac{ \sigma^{2}\alpha\sin[2\phi(h\alpha - d)] 
+4(h\alpha- d)\sinh \left( \frac{\sigma^2\phi  \alpha}{2}  \right)}{ \cosh \left(\frac{\sigma^2\phi 
\alpha}{2}   \right) +\cos[2\phi(h\alpha - d)]}\biggr\} =: g(\alpha,\phi).
\end{eqnarray}

The norm of the solution (\ref{felipeG}) is given by:

\begin{eqnarray}
R = \sqrt{2\pi} \; \sqrt[4]{\sigma^4+16h^2} \; e^{-\frac{\sigma^2}{8}(\alpha^2+\phi^2)}
\sqrt{\cosh \left(\frac{\sigma^2\phi\alpha}{2}\right) +\cos[2\phi(h\alpha - d)]}
\label{R}
\end{eqnarray}
\end{widetext}

Equations (\ref{gacomplex}),(\ref{gpcomplex}) give the directions of the geometrical tangents
to the trajectories which solves this planar system. We shall work in the gauge $N=e^{3\alpha}$.
By plotting the tangent direction field, it is possible to obtain the trajectories.
Due to the symmetries
$f(\alpha,\phi; h, d)=f(-\alpha,-\phi; -h, d)$, $g(\alpha,-\phi; h, d)=g(-\alpha,-\phi; -h, d)$,
and $f(\alpha,\phi; h, d)=-f(-\alpha,\phi; h, -d)$, $g(\alpha,\phi; h, d)=g(-\alpha,\phi; h, -d)$,
one concludes that a change in sign of $h$ corresponds to a inversion around the origin with time reversion, 
and a change in sign of $d$ corresponds to a reflexion in the $\phi$ axis. 
Hence, one can make definite choices of sign for these parameters without loss of generality.
 
\begin{figure}

\fbox{\includegraphics[height=75mm,width=75mm,]{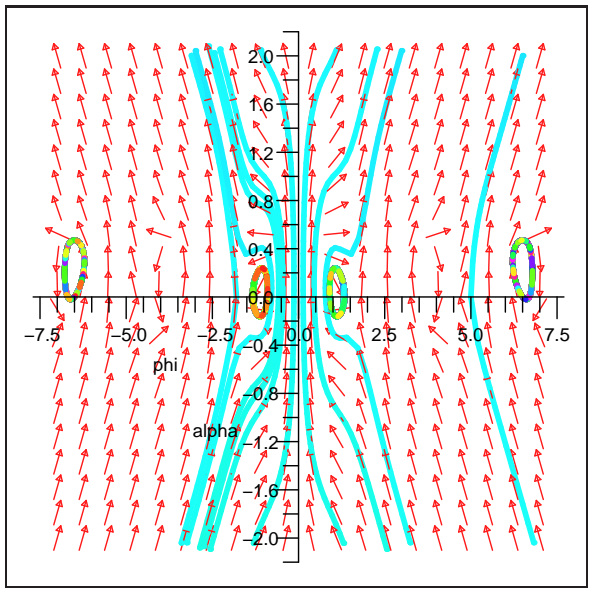}}

\caption{Field plot for the system (\ref{gacomplex}),(\ref{gpcomplex}), giving the direction of the geometrical tangent to the trajectories, for the values $\sigma^2=2, h=1/8$, and $d=-1$. }
\label{traj1}
\end{figure}

\begin{figure}

\fbox{\includegraphics[height=75mm,width=75mm,]{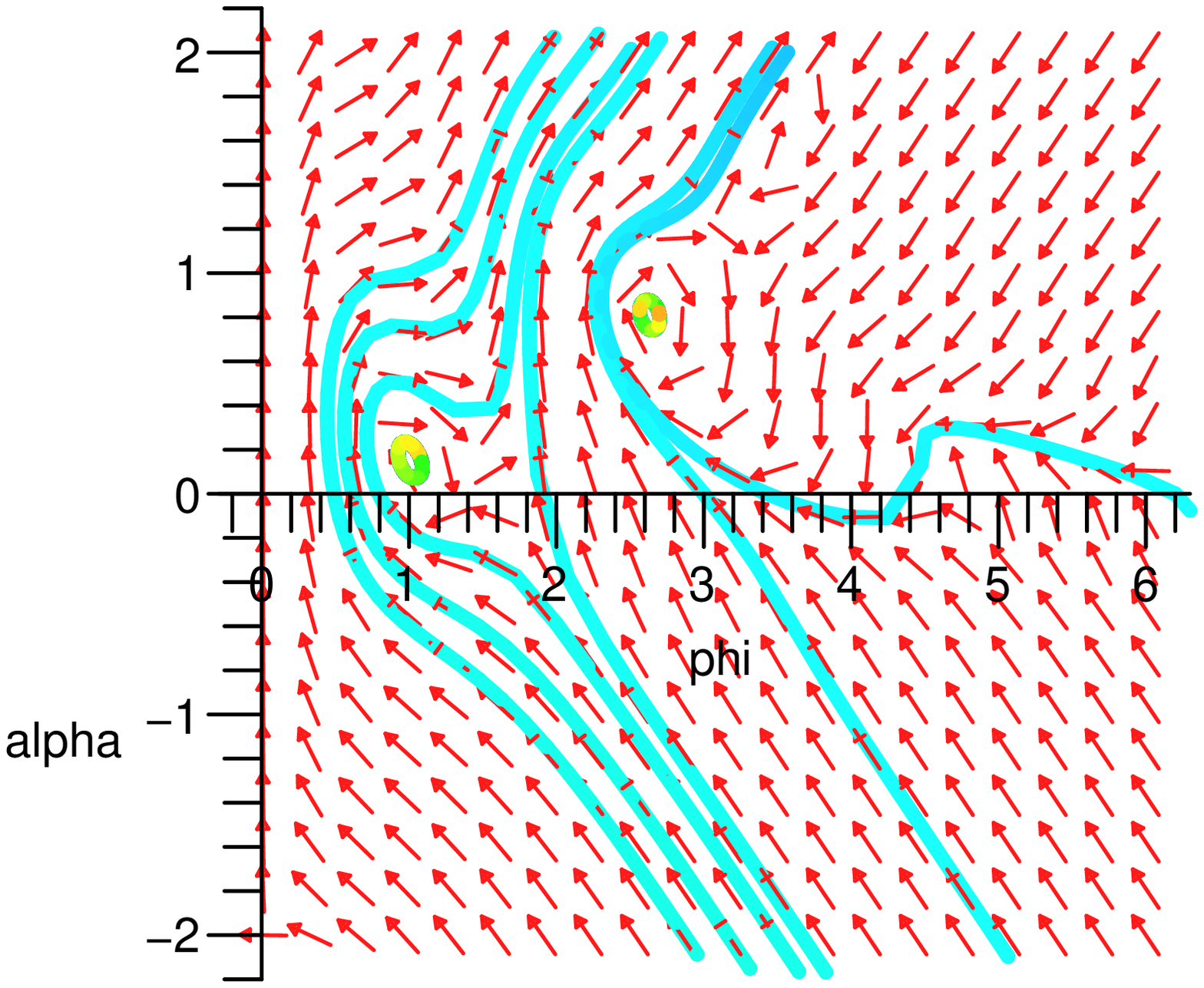}}

\caption{Field plot for the system (\ref{gacomplex}),(\ref{gpcomplex}), giving the direction of the geometrical tangent to the trajectories, for $\sigma^2=2, h=0.5,$ and $d=-1$. For a bigger value of $h$ the center points are farther from the axis $\alpha=0$. Note the change in the sign of $\dot{\alpha}$ when $\phi$ is big enough.}
\label{traj2}
\end{figure}

\begin{figure}

\fbox{\includegraphics[height=75mm,width=75mm,]{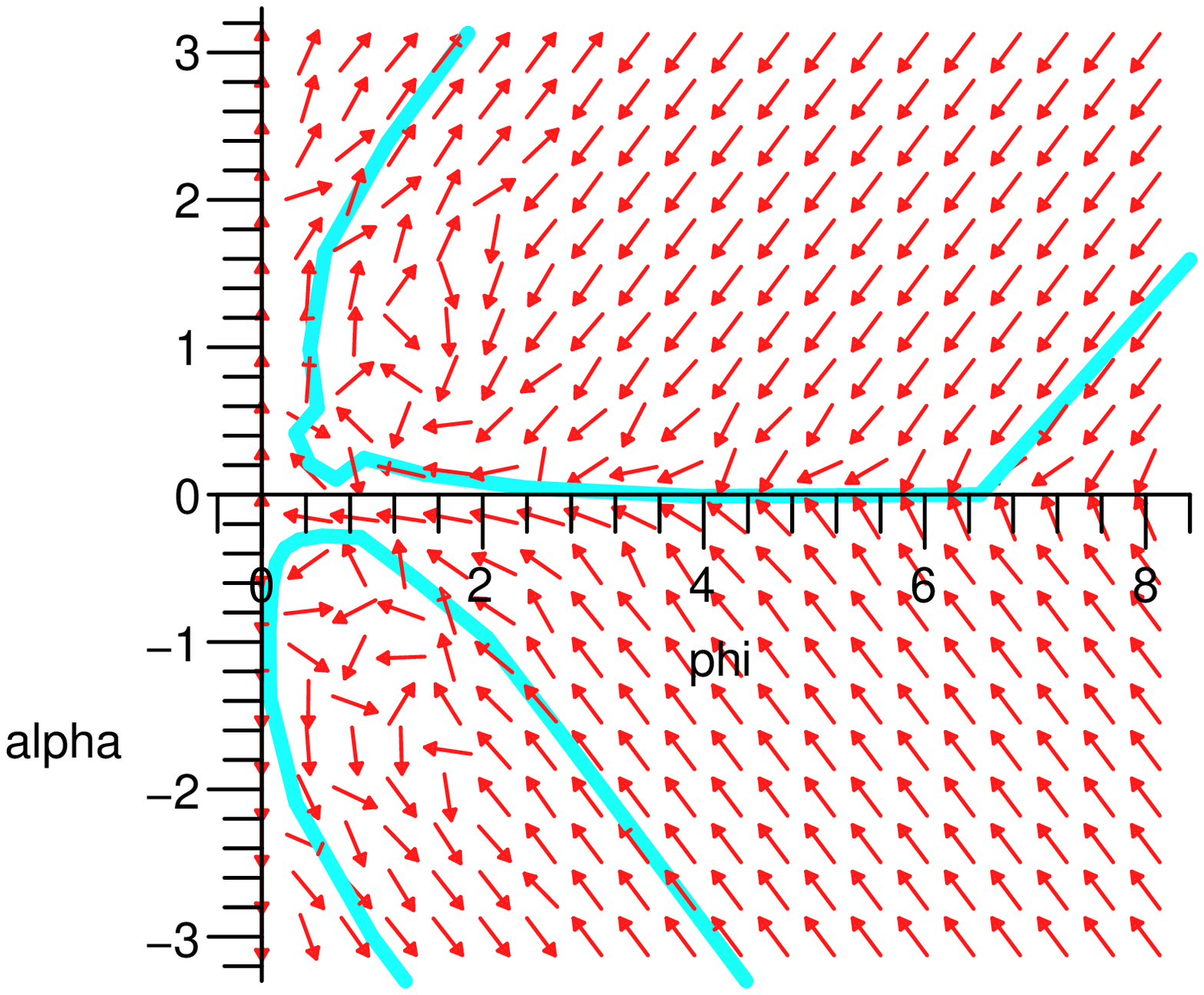}}

\caption{Field plot for the planar system (\ref{gacomplex}) (\ref{gpcomplex}) for  $\sigma^2=2, h=5$, and $d=-1$. Two trajectories are depicted: one representing a bouncing universe spending a long time on the bounce and other which corresponds to a universe which begins and ends in singular states (``big bang - big crunch'' universe). Note the change in the sign of $\dot{\alpha}$ when $\phi$ is big enough.}
\label{traj3}
\end{figure}

\begin{figure}

\fbox{\includegraphics[height=75mm,width=75mm,]{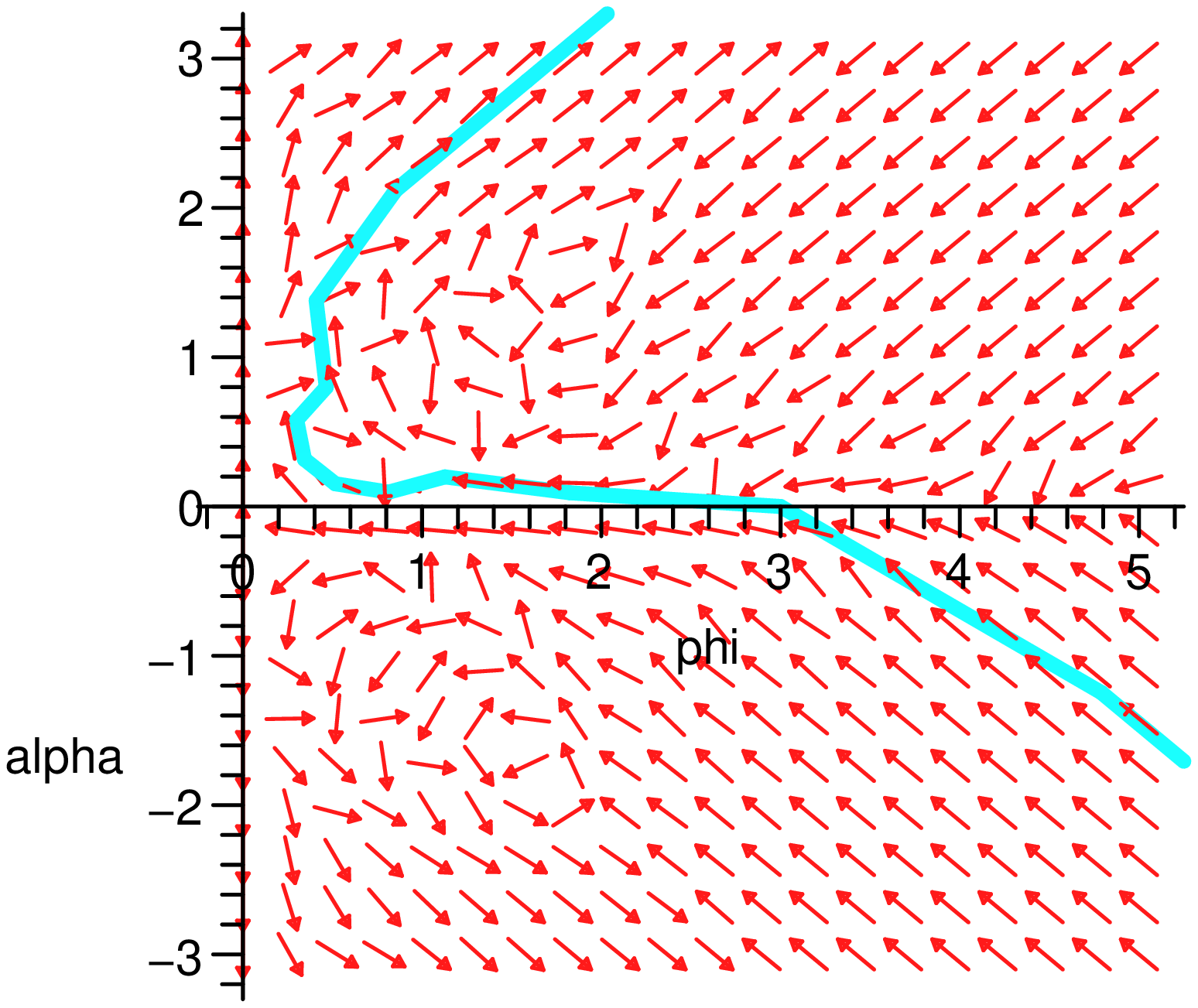}}

\caption{Field plot for the planar system (\ref{gacomplex}),(\ref{gpcomplex}) giving the direction of the geometrical tangent to the trajectories for $\sigma^2=2, h=5$, and $d=-1$. The trajectory for a universe coming from a singularity, experiencing a  long static phase and finally expanding, is depicted.}
\label{traj4}
\end{figure}

Field plots of this planar system are shown in Figs~(\ref{traj1},\ref{traj2},\ref{traj3},\ref{traj4}), 
for the choice of parameters $\{\sigma^2=2, h=1/8\}$, $\{\sigma^2=2, h=0.5\}$, and $\{\sigma^2=2, h=5\}$
(two portraits), respectively,
all with $d=-1$. 

The line $\phi = 0$ divides configuration space in two symmetric
regions as long as $f(\alpha,\phi)=f(\alpha,-\phi)$ in Eq.~(\ref{gacomplex}),
and $g(\alpha,\phi)=-g(\alpha,-\phi)$ in Eq.~(\ref{gpcomplex}). This can be seen in
Fig~(\ref{traj1}).

The line $\alpha = 0$ contains all the nodes of this
system, as it is shown in all figures. They appear when the
denominator of the above equations, which is proportional to the norm of
the wave function (see Eq.~(\ref{R})), is zero. No trajectory can pass through these points.
They happen when $\alpha = 0$ and $\cos (2 d \phi ) = -1$, or $\phi =
(2n+1)\pi /(2d)$, $n$ an integer, with separation $\pi/d$. 

The center
points appear when the numerators are zero, their locations depend on the
values of $h$, $\sigma^2$ and $d$, and they are not on the line $\alpha=0$,
unless $h=0$ (case of Refs.~\cite{fab2,santini}). Note in Fig~(\ref{traj1}), where 
$h$ is relatively much smaller than $\sigma^2$, that the centers are close but not
on the line $\alpha=0$. 

Apart the changes in the sign of 
$\dot{\alpha}$, which happen around the center points, there are regions with different
signs of $\dot{\alpha}$ when $|\phi|>>0, |\alpha|>>0$ where the hyperbolic functions dominate
over the trigonometric. For instance, considering $\phi>0$, if $d<0$,$h>0$, and $\alpha>0$ 
for $\alpha$ fixed, Eq.~(\ref{gacomplex}) shows that $\dot{\alpha}>0$ for small
$\phi$ when the $\cosh$ term dominates, but it change signs when $\phi$ is big enough in order
for the $\sinh$ term be greater than the $\cosh$ term in this equation, changing the sign of $\dot{\alpha}$.
This situation is depicted in Figs~(\ref{traj2},\ref{traj3},\ref{traj4}).

Note finally that the classical solutions for $\alpha(\tau)$ 
($a(\tau)\propto \tau^{1/3}$) and $\phi(\tau)$ are recovered when $\mid \alpha \mid \rightarrow
\infty$ or $\mid \phi \mid \rightarrow \infty$, and none of them are null. 

We can see plenty of different trajectories, depending on the
initial conditions and parameter values, in the 
Figs~(\ref{traj1},\ref{traj2},\ref{traj3},\ref{traj4}). 
Near the center points we can have oscillating
universes without singularities. One may have universes expanding classically from
a singularity, experiencing quantum effects in the middle of their
expansion, with possible accelerating and/or static phases, and recovering their classical
expansion behaviour for large values of $\alpha$. 

There are big bang-big crunch models, and
bouncing models, where the bounce may take long and connects two
asymptotically classical contracting and expanding phases
(see Fig (\ref{traj3}). 

Note that, if we choose $V(k)=A(-k)$ (see Eqs  (\ref{sol0k},\ref{gauss1})), 
one obtains the same field plots as above with the axis
$\alpha$ and $\phi$ interchanged, obtaining
more possibilities for bouncing models. 

Finally, there is the special situation of $\sigma^{2}=0$, which leads
to qualitative different solutions, as we will see in the next section.

\section{non-singular inflationary Bohmian trajectories}

There is an interesting case which is obtained when $\sigma^{2}=0$ in the gaussian (\ref{gauss}).

Hence, $A(k)$ reads
\begin{equation}
\label{ak} 
A(k) = \exp\biggr[i\frac{(k-\sqrt{2}d)^2}{4 h}\biggl] \quad,
\end{equation}
Then the wave function (\ref{felipeG}) reduces to
\vspace{0.5cm}

\begin{eqnarray}
\label{felipe}
\Psi(u,v) &=& 2\sqrt{\pi |h|} \biggl[\exp i\biggl(-h u^2 + \sqrt{2}du 
+ \frac{\pi}{4}\biggr)\nonumber \\ &+&
\exp i\biggl(-h v^2+ \sqrt{2}dv + \frac{\pi}{4}\biggr)\biggr] \quad.
\end{eqnarray}

Its norm is given by $R= 4 \sqrt{\pi |h|}\cos[\phi(h\alpha-d)]$, yielding the quantum potential
\begin{equation}
\label{qfelipe} 
Q=  (h \alpha - d)^2-h^2\phi^2 \quad.
\end{equation}

The guidance relations given by (\ref{gacomplex}) and (\ref{gpcomplex}) now reduce to
\begin{equation}
\label{guialphaf}
\dot{\alpha}=h\alpha-d\quad ,
\end{equation}
\begin{equation}
\label{guiphif}
\dot{\phi}=-h\phi\quad .
\end{equation}
The equations (\ref{guialphaf}) and (\ref{guiphif}) represent a linear dynamical system. 
The only critical point $(\phi=0, \alpha=\frac{d}{h})$ is a saddle point and, as it is well known, 
it represents an unstable equilibrium.

\begin{figure}

\fbox{\includegraphics[height=75mm,width=75mm,]{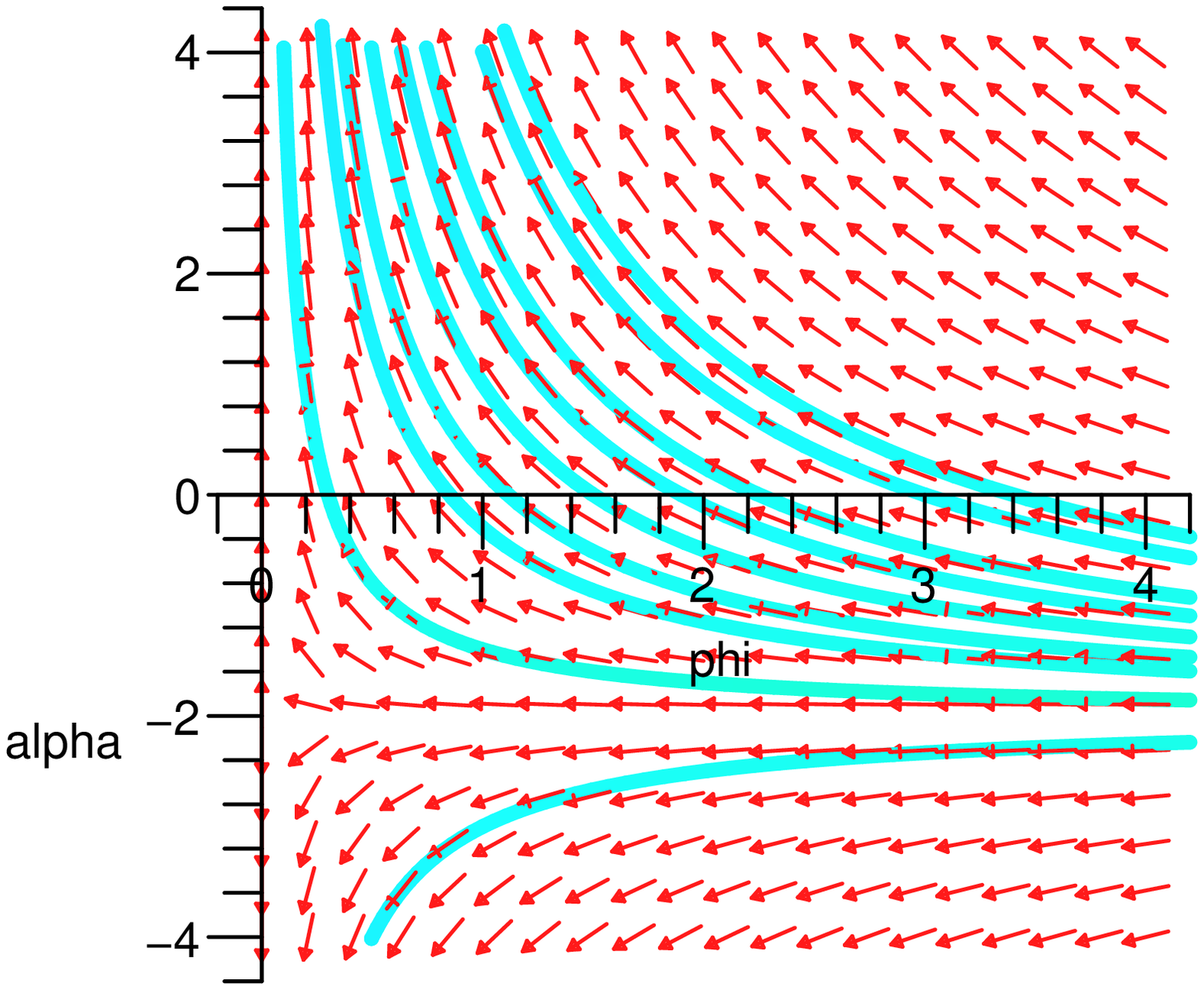}}

\caption{Field plot for the linear planar system (\ref{guialphaf}) (\ref{guiphif}) for   $\sigma^2=0, h=0.5$, and $d=-1$. A class of trajectories represent an universe that begins to inflate quantum mechanically from a Planckian size closed flat space-time in the infinite past, and when it becomes large enough it undergoes an analytical graceful exit to a decelerated classical stiff matter expansion phase.}
\label{traj5}
\end{figure}

The field plot of these solutions is depicted in Fig~(\ref{traj5}) for $d=-1, h=0.5$. Note
that there are two regions of different signs of $\dot{\alpha}$ separated by the line
$\alpha=d/h$.

In this case there are analytical solutions, which read
\begin{equation}
\label{solf}
a=e^{\alpha}=e^{d/h}\exp (\alpha_0 e^{ht})\;\; {\rm and} \;\; \phi=\phi_0 e^{-ht}\quad ,
\end{equation}
where $\alpha_0$ and $\phi_0$ are integration constants, and 
remembering that the time parameter $t$ is related to cosmic time $\tau$ 
through $\tau = \int dt e^{3\alpha(t)}\Rightarrow \tau-\tau_0 = {\rm Ei}(3\alpha_0e^{ht})/h$,
where Ei$(x)$ is the exponential-integral function.

These solutions represent ever expanding or contracting non-singular models, depending on the sign of $h$.
Let us consider the physical situation of expanding solutions with $h>0$. The Hubble and deceleration
parameters $a'/a$ and $a''/a$ read (a prime denotes a derivative in cosmic time)
\begin{equation}
\label{hub}
\frac{a'}{a}=\frac{\alpha_0 h e^{ht}}{a^3} \quad ,
\end{equation}
and
\begin{equation}
\label{dec}
\frac{a''}{a}=\frac{\alpha_0 h^2}{a^6}e^{ht}(1-2\alpha_0 e^{ht})\quad ,
\end{equation}
and the scalar curvature is
\begin{equation}
\label{Rscalar} 
{\cal R} = -\frac{6\alpha_0 h^2}{a^6}e^{ht}(1-\alpha_0 e^{ht})\quad .
\end{equation}

There are three important phases in this model. For $t<<0$ the Universe expands accelerately 
from its minimum size $a_0= e^{d/h}$ (remember that for the physical scale factor one
has $a_0^{\rm phys} = l e^{d/h}/\sqrt{2V}$), which occurs in the infinity past 
$t\rightarrow-\infty$ when the curvature is null but increasing while scale factor grows. 
The scalar field is very large in that phase.
For $t>>0$ the Universe expands decelerately, the scale factor is immensely big, the scalar field
becomes negligible and the curvature approaches zero again. The transition occurs when
$ht_{{\rm tran}}=-\ln (2\alpha_0)$. 

Around $ht=0$ one has
\begin{equation}
a\approx e^{\alpha_0+d/h}[1+\alpha_0 ht + (\alpha_0 h^2 + \alpha_0^2h^2)t^2/2!+...].
\end{equation}
If $\alpha_0 >> 1$ (and hence $t>t_{{\rm tran}}$, which means in the deceleration phase), 
one can write $a\approx e^{\alpha_0+d/h}\exp (\alpha_0 h t)$. In that case, from 
$\tau = \int dt a^3(t)$, one obtains that $a\propto (\tau-\tau_0)^{1/3}$ and 
$\phi '\propto 1/\tau \propto 1/a^3$, as in the classical regime. 

Collecting all these phases, and considering $\alpha_0 >>1$, 
one has a non-singular ever expanding Universe,
starting with a constant and finite size (which may be of the order of
the Planck length if $d=0$, a Planckian flat space) in the infinity past, 
which inflates afterwards till it attains an almost \footnote{We would like to say that it is not necessary that classical
behaviour appears
when the Universe is large, see \cite{HH}.
In some circumstances, a quantum behavior should be desirable, as in \cite{santini} to produce late
acceleration
in the Universe, or to avoid a Big Rip, see \cite{bigrip} .}classical decelerating
expanding regime, with a size $e^{\alpha_0}$ times bigger than it was 
initially.  At this time, radiation may start to dominate over the scalar field
since it becomes much smaller than in the infinity past, and its energy 
density goes may like $1/a^6$. There are no event nor particle horizons.
Hence we have an inflationary non-singular model which
can be smoothly joined to the standard model.

Some types of bouncing model were obtained in \cite{fab2}. 
In this reference, the matter content of the primordial model was
considered to be stiff matter modeled by a free massless scalar field minimally
coupled to gravity. The Wheeler-DeWitt equation turns out to be a
two-dimensional Klein-Gordon equation, and gaussian superpositions, 
with positive width, of negative and
positive frequency modes solutions were considered. In the flat case, a
two dimensional dynamical system for the bohmian trajectories was obtained, yielding
a variety of possibilities: big bang-big crunch models, oscillating universes, 
bouncing solutions, and ever expanding (contracting) big bang (big crunch) models with a period of acceleration
in the middle of their evolutions. No non-singular purely expanding model with a primordial
accelerated phase (a non-singular inflationary model) was obtained (models of this type
were obtained only with a non-minimal coupling between the scalar and gravitational fields,
\cite{nonminimal}). 

\section{Conclusion}

In the present paper we generalized the gaussians superpositions of Refs.~\cite{fab2}\cite{santini} to gaussians with complex widths with non negative real part. We obtained a richer two dimensional dynamical system, with oscillating
universes without singularities, universes expanding classically from
a singularity, experiencing quantum effects in the middle of their
expansion with possible accelerating and/or static phases whose duration
depend on the three free parameters of the gaussian (its complex width and the
location of its center) and two initial conditions, which recover their classical
expansion behaviour for large scale factors. There are also big bang-big crunch models, and
bouncing models, where the bounce may take long, and connects two
asymptotically classical contracting and expanding phases. 

The most interesting solutions occur when the real part of the width is made zero. Then it was obtained analytical non-singular inflationary models which expand accelerately in the infinity past from flat universes with finite (not zero) size spatial sections, and are smoothly connected to a classical decelerated stiff matter expanding phase. It is like the pre-big bang model \cite{pbb}, with a minimum volume spatial section in the infinity past (which can be of the order of the Plank volume), or as the emergent model \cite{ellis} for flat spatial sections, without any gracefull exit problem.
Sufficient inflation can be obtained with reasonable choices of initial conditions ($\alpha_0 > 70$). The picture is then of a Universe that begins to inflate quantum mechanically from a Planckian size closed flat space-time in the infinity past,
and when it becomes large enough it makes a smooth transition to a decelerated classical
stiff matter expansion, which will be eventually be dominated by radiation before nucleosynthesis.
Some reheating process must then also be presented in our model,
and this could happen through interactions of the type $\sqrt{-g}\upsilon\phi\Psi\bar{\Psi}$,
where $\Psi$ represents some fermion field, $\upsilon$ is a coupling constant,
and the scalar field decays into these fermions. Such fermions would
not be present in the beginning but they would have been produced when the universe
got bigger and transitated to the decelerated phase. Such extra term  
would not be relevant in the hamiltonian which describes the model
when the universe was small (it is proprtional to $a^3$) and empty of these fermions,
justifying our quantum solutions in this regime. However, when the fermions begin to be
produced afterwards, our quantum solutions could not be anymore reliable.
At this stage, when the model becomes decelerated and gets closer to classical
behaviour, this solution must be connected to the classical radiation plus
stiff matter solution . The nonclassical behaviour of our bohmian quantum solution for 
$t > 0$ should then be of course discarded, they are out of the range of the validity of our assumptions.
Therefore, our non-singular inflationary solution should be
utilized only as a zeroth-order approximation to a realistic cosmological
inflationary model, in the same way as the de Sitter model or power law solutions are used for usual inflation \cite{mukhanov-book}. The improvement of the model, including reheating and a radiation field deserves further research as there are many possibilities to be investigated. Another
important step forward is to evaluate the evolution of cosmological
perturbations using the formalism developped in Ref.~\cite{emanuel}, and confront it with
observations \cite{wmap}. This will be the subject of our future
publications.

\section*{ACKNOWLEDGEMENTS}

Two of us (FTF and NPN)  would like to thank the French/Brazillian cooperation
CAPES/COFECUB for partial financial support.
One of us (NPN) would like to thank CNPq of Brazil for financial support. ESS would like to thank CGMI/CNEN-MCT for technical support.
We would also like to thank `Pequeno Seminario' of CBPF's Cosmology Group
for useful discussions, specially Prof. Santiago P\'erez Bergliaffa for his comments and 
suggestions.


\begin{thebibliography}{99}

\bibitem{inflation} A.~Guth, \prd {\bf 23}, 347 (1981); A.~Linde,
Phys.  Lett. B {\bf 108},389 (1982); A.~Albrecht and P.~J.~Steinhardt,
\prl {\bf 48}, 1220 (1982); A.~Linde, Phys. Lett. B {\bf 129}, 177
(1983); A.~A.~Starobinsky, Pis'ma Zh. Eksp.  Teor. Fiz. {\bf 30}, 719
(1979) [JETP Lett. {\bf 30}, 682 (1979)]; S.~Hawking, Phys. Lett. B {\bf 115},
295 (1982); A.~A.~Starobinsky, Phys. Lett. B {\bf 117}, 175 (1982);
J.~M.~Bardeen, P.~J.~Steinhardt, and M.~S.~Turner, \prd {\bf 28}, 679
(1983); A.~Guth, S.~Y.~Pi, \prl {\bf 49}, 1110 (1982).

\bibitem{muk}  V.~Mukhanov and G.~Chibisov,
JETP Lett. {\bf 33}, 532 (1981).

\bibitem{wmap} D.~N.~Spergel \etal, Ap. J. Suppl. {\bf 148}, 175
(2003); D.~N.~Spergel \etal, {\tt astro-ph/0603449}, Ap. J. (2006).

\bibitem{transP} J.~Martin and R.~Brandenberger, \prd {\bf 63}, 123501
(2001); R.~Brandenberger and J.~Martin, Mod. Phys. Lett. A {\bf 16},
999 (2001); J.~Niemeyer, \prd {\bf 63}, 123502 (2001); M.~Lemoine,
M.~Lubo, J.~Martin, and J.~P.~Uzan, \prd {\bf 65}, 023510 (2002).

\bibitem{singularity} S.~W.~Hawking and G.~F.~R.~Ellis, {\it The large
scale structure of space-time}, Cambridge University Press (1973);
R.~M.~Wald., {\it General Relativity}, Chicago University Press
(1984). See also A.~Borde and A.~Vilenkin, \prd {\bf 56}, 717 (1997)
for a more recent discussion.

\bibitem{bohm} D. Bohm, Phys. Rev. {\bf 85}, 166 (1952);
D. Bohm, B. J. Hiley and P. N. Kaloyerou, Phys. Rep. {\bf 144}, 349
(1987).

\bibitem{holland} P. R. Holland, {\it The Quantum Theory of Motion: An
Account of the de Broglie-Bohm Causal Interpretation of Quantum
Mechanichs} (Cambridge University Press, Cambridge, 1993).

\bibitem{santini0} N. Pinto-Neto and E. Sergio Santini, Phys.Rev. {D \bf 59}
123517 (1999).

\bibitem{cons} N. Pinto-Neto and E. Sergio Santini, Gen. Rel. and Grav. {\bf 34},
505 (2002).

\bibitem{bola} J. C. Vink, Nucl. Phys. {\bf B369}, 707 (1992); 
Y. V. Shtanov,  Phys. Rev. {\bf D54}, 2564 (1996); 
A. Valentini, Phys. Lett. {\bf A158}, 1, (1991); 
J. A. de Barros and N. Pinto-Neto,
Int. J. of Mod. Phys. {\bf D7}, 201 (1998).

\bibitem{many} {\it The Many-Worlds Interpretation of
Quantum Mechanics}, ed. by B. S. DeWitt and N. Graham (Princeton
University Press, Princeton, 1973).

\bibitem{kuc} K. Kuchar in {\it Quantum Gravity 2: A Second Oxford
Symposium}, eds. C. J. Isham, R. Penrose and D. W. Sciama
(Clarendon Press, Oxford, 1981).

\bibitem{har} R. B. Griffiths, Journal of St. Phys. {\bf 36}, 219 (1984);
M. Gell-Mann and J. B. Hartle in {\it Complexity, Entropy
and the Physics of Information}, ed. by W. Zurek (Addison Wesley, 1990);
J. B. Hartle in the {\it Proceedings of the $13^{th}$
International Conference on General Relativity and Gravitation},
ed. by R. J. Gleiser, C. N. Kozameh and O. M. Moreschi (Institute
of Physics Publishing, London, 1993); 
R. Omn\`es, {\it The Interpretation of Quantum Mechanics}
(Princeton University Press, Princeton, 1994).

\bibitem{banks} T. Banks, Nucl. Phys. {\bf B249}, 332 (1985).

\bibitem{pad} T. P. Singh and T. Padmanabhan, Ann. Phys.
{\bf 196}, 296 (1989).

\bibitem{kie} D. Giulini and C. Kiefer, Class. Quantum Grav. {\bf 12}, 
403 (1995).

\bibitem{hal} J.J. Halliwell, in {\it Quantum Cosmology and Baby
Universes}, ed. by S. Coleman, J.B. Hartle, T. Piran and S. Weinberg
(World Scientific, Singapore, 1991).

\bibitem{bounce} J.~Acacio de Barros, N.~Pinto-Neto, and
M.~A.~Sagioro-Leal, Phys. Lett. A {\bf 241}, 229 (1998);
F.G. Alvarenga, J.C. Fabris, N.A. Lemos and G.A. Monerat,
Gen.Rel.Grav. {\bf 34}, 651 (2002); 
N. Pinto-Neto, E. Sergio Santini and Felipe T. Falciano, Phys. Lett.{\bf  A 344}, 131-143(2005).

\bibitem{loop} M. Bojowald, Phys. Rev. Lett. {\bf 86}, 5227 (2001);
M. Bojowald, Phys. Rev. D {\bf 64}, 084018 (2001); 
A. Ashtekar, M. Bojowald and J. Lewandowski Adv. Theor. Math. Phys. {\bf 7}, 233 (2003).

\bibitem{emanuel} P. Peter, E. Pinho and N. Pinto-Neto, JCAP {\bf 07}, 014 (2005);
P. Peter, E. Pinho and N. Pinto-Neto, Phys. Rev. {\bf D73}, 104017 (2006);
P. Peter, E. Pinho and N. Pinto-Neto, Phys.Rev. {\bf D75}, 023516 (2007); 
E. Pinho and NPN, hep-th/0610192, to appear in Phys. Rev. D. 

\bibitem{fab2} R.~Colistete Jr., J.~C.~Fabris, and N.~Pinto-Neto,
\prd {\bf 62}, 083507 (2000).

\bibitem{santini} N Pinto-Neto and  E. Sergio Santini, Phys. Lett. {\bf A 315}, 36 (2003).

\bibitem{HH} J. Halliwell and J. Hartle, Phys. Rev. {\bf D41}, 1815 (1990).

\bibitem{bigrip} Mariusz P. Dabrowski, Claus Kiefer, Barbara Sandhoefer, Phys.Rev. {\bf D74} 044022 (2006).

\bibitem{zeldovich} Ya. B. Zel'dovich, Sov. Phys. - JETP {\bf 14} 1143 (1962).

\bibitem{nonminimal} N. Pinto-Neto and R. Colistete Jr,
Phys. Lett. {\bf A290}, 219 (2001).

\bibitem{pbb} G.~Veneziano, Phys. Lett. B {\bf 265}, 287 (1991);
M.~Gasperini and G.~Veneziano, Astropart. Phys. {\bf 1}, 317 (1993);
See also J.~E.~Lidsey, D.~Wands, and E.~J.~Copeland, Phys. Rep. {\bf
337}, 343 (2000) and G.~Veneziano, in {\sl The primordial Universe},
Les Houches, session LXXI, edited by P.~Bin\'etruy {\it et al.}, (EDP
Science \& Springer, Paris, 2000).

\bibitem{ellis} David J. Mulryne, Reza Tavakol, James E. Lidsey and George F. R. Ellis,
Phys. Rev. {\bf D71}, 123512 (2005). 

\bibitem{top} J. Martin, N. Pinto-Neto and I. D. Soares,
JHEP {\bf 0503}, 060 (2005).

\bibitem{bola27} J. A. de Barros and N. Pinto-Neto,
Int. J. of Mod. Phys. {\bf D7}, 201 (1998).

\bibitem{tese} E. Sergio Santini, PhD Thesis, CBPF-Rio de Janeiro, (May 2000),
(gr-qc/0005092).

\bibitem{mukhanov-book} V. Mukhanov,  {\it Physical Foundations of Cosmology}, (Cambridge University Press, Cambridge, 2006).

\end{thebibliography}
\end{document}